\documentclass[11pt]{article}
\usepackage[dvips]{graphicx}
 \newcommand{\cc}{\cite}

\newcommand{\be}{\begin{equation}}
\newcommand{\ee}{\end{equation}}

\def\lagr{\hbox{$\cal L$}}
\def\ve{\varepsilon}
\def\w{\omega}

\def\pd{\partial}

\def\L{\Lambda}

\def\W{\Omega}
\def\z{\zeta}

\def\<{\langle}
\def\>{\rangle}

\def\Log{\hbox{ln}}

\def\g{\gamma}  
  
   \def\L{\Lambda}

\def\m{\mu}

\def\z{\zeta}
\def\w{\omega}
\def\tt{\theta}
\def\v{\vec}

\def\({\left(}
\def\[{\left[}
\def\){\right)}
\def\]{\right]}

\def\pd{\partial}

\def\w1{W^{(1)}}
\def\v1{V^{(1)}}

\def\lagr{\hbox{$\cal L$}}
\def\ve{\varepsilon}
\def\w{\omega}

\def\pd{\partial}

\def\L{\Lambda}

\def\W{\Omega}
\def\z{\zeta}

\def\<{\langle}
\def\>{\rangle}

\def\Log{\hbox{ln}}

\def\g{\gamma}  
  
   \def\L{\Lambda}

\def\m{\mu}

\def\z{\zeta}
\def\w{\omega}
\def\tt{\theta}
\def\v{\vec}

\def\({\left(}
\def\[{\left[}
\def\){\right)}
\def\]{\right]}

\begin{document}
 \begin{center}
 {\large \bf{Renormalization Group in Zero-Point Energy Calculations}} \\
   \vspace{5mm}
 {\small Igor O. Cherednikov$^{a,b,}$\footnote{Igor.Cherednikov@jinr.ru}}
\\
 \vspace{2mm}
 $^a$ {\small \it Bogolyubov Laboratory of Theoretical Physics,
Joint Institute for Nuclear Research \\
141980 Dubna, Russia}\\
\vspace{2mm}
$^b$ {\small \it Institute for Theoretical Problems of Microphysics,
Moscow State University \\ 119899 Moscow, Russia}
\end{center}
\date{ }
\begin{abstract}
\noindent
{\small The simple consequences of the renormalization group invariance in calculations of the ground state
energy for models of confined quantum fields are discussed. The case of (1+1)D MIT quark bag
model is considered in detail. }
\end{abstract}

\noindent
Recently, the application of the renormalization group methods in study of the zero-point energy
effects for various situations attracted an attention of community \cc{1}.
The calculations of zero-point energy for quantized fields under nontrivial
boundary conditions encounter a number of difficulties (for the
most recent review on these problems see \cc{2}). A majority of them
are connected with ambiguities in results obtained by means of different
regularization and renormalization methods. One of the physically interesting
problems is the dependence of the (renormalized) energy on additional mass
parameter, which emerges inevitably in any regularization scheme. For
example, in the widely used $\z$-function regularization approach
the one-loop vacuum energy for a fermion field may be defined as
\be
\ve_f = - K_D
\(\z'_f(0) - \z_f(0)\Log{m \over \m}\) \ , \label{zf}
\ee where $K_D$ is the constant depending on the space-time dimension, $m$ is a field mass,
and the arbitrary mass $\m$ must
be introduced in order to restore the correct dimension of the corresponding
zeta-functions \cc{3}.
Within any renormalization procedure, the finite part of the energy would
contain, in general, a $\m$-dependent contribution. Of course, there are
several situations, for which $\z (0) = 0$ and  this dependence is obviously
cancelled due to some geometrical, or other, properties of the given configuration.
However, it would be useful to investigate more general case.

In this talk, the renormalization of zero-point
energy is considered from the point of view of the convenient quantum field theory, what means that
any variations of the mass scale $\m$ must produce no changes of the physically
observable quantities. This requirement naturally leads to a sort of the
renormalization group equation, the solution of which allows to conclude that
some of the parameters of the ``classical'' mass formula have to be considered
as running constants. This may be important, { \it e. g.}, in some
phenomenological applications, such as the quark bag models where the zero-point energy yields
a nontrivial contribution to the mass of hadron, since it may
provide a deeper understanding  of relations between fundamental and effective
theories of the hadronic structure.

For our purposes, it would be enough to use the following form of the $\z$-regularized ground
state energy of the confined fermion field:
\be E = -
\sum \w_n \to E_{reg} (\m, \ve) = -\m^{\ve} \sum \w^{1-\ve}_n \ . \ee
The presence of a mass may lead, in general,  to some new effects compared to the massless
case, and contains additional divergences that
have to be subtracted. Let us consider the (1+1)D MIT bag model with the
massive fermions in more detail \cc{4}. The Lagrangian of this system
\be
\lagr_{MIT} = i\bar\psi\g\pd\psi - \tt(|x|<R) \ \(m \bar\psi\psi + B\)  -
\tt(|x|>R)\ M \bar\psi\psi   \ee
describes (in the limit $M\to\infty$) the fermion field confined to the segment $[-R,R]$
under the (1+1)D boundary condition:
\be
(\pm i \g^1 +1)\psi(\pm R) = 0\ .
\ee
The exact fermion energy spectrum reads:
\be
\w_n = \sqrt{\({\pi \over 2R} n + {\pi \over 4R}\)^2 + m^2} \ .
\ee
In this simple case it is not necessary to use the heat-kernel expansion (see, {\it e.g.},
\cc{2}. \cc{3}) since the spectrum of eigenvalues is known explicitly.
Here we will be interesting only in the small mass $m$ limit, so we drop out
all terms of the order $m^4$ and higher, what corresponds to the expansion in vicinity of the
chiral limit \cc{4}. Then the eigenvalues $\w_n$ can be
written as
\be
\w_n = \W_1 n + \W_0 + {m^2 \over 2\(\W_1 n +\W_0\)} + O(m^4) \ ,
\ee
where
\be
\W_1 = {\pi \over 2R} \ \ , \ \ \W_0 = {\pi \over 4R} \ .
\ee
In order to analyze the singularities in the ground state energy, we use the
expansion for $n > 0$:
\be
\w_n = \W_1 n + \W_0 + {\W_{-1} \over n} + O(n^{-2}) \ ,
\ee
where $\W_{-1} = m^2 R/\pi$, and assume the lowest quark state with $\w_0 =
\W_0 + 2\W_{-1}$ to be filled.

It can be shown, that the $\z$-regularized sum (1) reads:
\be
E_{reg} = -\W_{-1} \({1 \over \ve} + \g_E \) + {\W_1 \over 12} + {\W_0 \over
2} + {\W_0^2 \over 2 \W_1} + \W_{-1}\(\Log {\W_1 \over \m} +1 \) , \label{div}
\ee where $\g_E = 0.5772...$ is the Euler constant.
It's interesting to note, that the regularization by the exponential cutoff
gives the equivalent result up to additional power-law divergences \cc{5, 6}. These divergences are
a generic feature of the regularization schemes which use an UV cutoff, and, in contrast,
never emerge in schemes without it, such as dimensional and $\z$-function regularization.

The divergent part of (\ref{div}) can be extracted in the form:
\be
E_{div} = - {m^2 R \over \pi} \({1 \over \ve} + \g_E - \Log {\pi \over 8} \ .
\label{ms} \)
\ee
We include in $E_{div}$ the pole $\ve^{-1}$  as well as the
transcendent numbers $\g_E$ and $\Log {\pi \over 8}$ in analogy to the widely
used scheme $\overline{\hbox{MS}}$ in QFT, but we should mention that this analogy is
only formal one, since (\ref{ms}) has
nothing to do with the singularities appearing in the conventional field
theory since it depends on the geometrical parameter $R$.
The removing of divergent part in (\ref{div}) is performed by the absorption of $E_{div}$
(\ref{ms}) into the definition of the ``classical'' bag constant $B$ \cc{5}, which is
introduced in the mass formula and characterizes the energy excess inside the
bag volume as compared to the energy of nonperturbative vacuum outside \cc{4}.

The finite energy of our bag with one fermion on the
lowest energy level is
\be
E(R, \m) = 2B_0 R + {11 \pi \over 48 R} + {3 m^2 R \over \pi} - {m^2 R \over
\pi } \Log \m R  \ , \ee where $B_0$ is the renormalized bag constant.
This quantity, being a physical observable, should not depend on a choice of
arbitrary scale $\m$. Then the condition of the renormalization invariance has to be
imposed on it, what yields
\be
\m {d \over d\m} E(B_0(\m), \m) = 2 R \g_B - {m^2 R \over \pi} = 0 \ ,
\label{cond1}
\ee where
\be
\g_B = \m {d \over d\m} B_0(\m) \ .
\ee
The solution of this equation can be written as
\be
B_0 (\m) = B_0 (\m_0) + {m^2 \over 2 \pi} \Log {\m \over \m_0} \ ,
\label{run1}
\ee where the value $B_0(\m_0)$ gives the boundary condition for the
solution of the differential equation (\ref{cond1}). It may seem that the
running parameter $B_0(\m)$ depend on both the initial value $B_0(\m_0)$ and
$\m_0$ itself, but in fact it must not depend on the starting point. Then
it's convenient to express the running constant in terms of a single
variable (thus the Eq. (\ref{cond1}) allows to exclude an extra parameter):
\be
\L_{MIT} = \m_0 \exp\[-{2\pi \over m^2}B_0(\m_0)\] \ , \label{lam1}
\ee and write
\be
B_0(\m) = {m^2 \over 2 \pi} \Log {\m \over \L_{MIT}} \ . \label{lam2}
\ee
The scale $\L_{MIT}$ appears to be an analogue of the fundamental scale $\L_{QCD}$ in
Quantum Chromodynamics. One can easily check that the total bag energy $E(B_0(\m), \m)$ is
independent on choice of the points $\m$ and $\m_0$.

The actual size of the bag can be found from the equation which determines
the energy minimum:
\be
{\pd E \over \pd R} = 0 \ . \label{mr1}
\ee
Then, the radius $R_0$ obeys the relation
\be
{11 \pi \over 48 R_0^2} = 2\(B_0 + {m^2 \over \pi}\) - {m^2 \over \pi} \Log \m
R_0  \ .  \label{mr2}
\ee
Taking into account the formula (\ref{lam2}) for the running constant
$B_0(\m)$, one finds the following relation between the fundamental scale
$\L_{MIT}$ and the bag's radius:
\be
\L_{MIT} = R_0^{-1} \exp\[2-{11 \over 48} \({\pi \over m R_0}\)^2\] \ .
\ee
Therefore, we find that in the RG-improved version of the (1+1)D MIT bag
model, which takes into account the renormalized fermion ground state
energy, the ``bag constant'' should be considered as a running parameter,
and the value of the bag energy as well as it's radius are determined by the
single (except of the current quark mass $m$) dimensional parameter
$\L_{MIT}$ which plays a role of the fundamental energy scale, similar to
$\L_{QCD}$, and should also be determined experimentally.
Note, that the relation (\ref{lam1}) can be considered as an
improvement of the straightforward identification of the scale $\m$ and
$\L_{MIT}$ proposed earlier.
It is clear, that in more complicated situations, for example, in (3+1)D MIT bag model with a
larger amount of coupling constants, the results of the RG analysis would be rather nontrivial.

We have considered one of the simplest the consequences of the renormalization invariance condition
in the zero-point energy evaluation in the quark bag models.
It is shown that the value of the bag mass is
controlled by the fundamental energy scale analogous to $\L_{QCD}$, while
the ``bag constant'' becomes a running parameter. The effects of the
renormalization invariance in more realistic models -- such as (3+1)D chiral hybrid quark
bags, will be studied elsewhere.

The author is most grateful to the Organizers of the Conference RG2002
for the warm hospitality and financial support. This work is supported also
by RFBR (projects nos. 02-02-16194, 01-02-16431 and 00-15-96577), and INTAS (project
no. 00-00-366).


\end{document}